# DESIGN AND HIGH POWER TESTING OF 52.809 MHZ RF CAVITIES FOR SLIP STACKING IN THE FERMILAB RECYCLER*

R. Madrak#, D. Wildman, FNAL, Batavia, IL 60510, USA

## Abstract

For NOvA and future experiments requiring high intensity proton beams, Fermilab is in the process of upgrading the existing accelerator complex for increased proton production. One such improvement is to reduce the Main Injector cycle time, by performing slip stacking, previously done in the Main Injector, in the now repurposed Recycler Ring. Recycler slip stacking requires two new RF cavities operating at slightly different frequencies (df = 1260Hz). These are copper, coaxial, quarter wave cavities with R/Q =13 ohms. They operate at a peak gap voltage of 150 kV with 150 kW peak drive power (60% duty factor), and are resonant at 52.809 MHz with a 10 kHz tuning range. Two have been completed and installed. The design, high power test results, and status of the cavities are presented.

## INTRODUCTION

To increase beam power, the Recycler Ring has been converted from an antiproton storage ring to a proton preinjector for proton injection/slip stacking. The increase in power, which is projected to be 80%, is due to a shorter Main Injector cycle time (decreased from 2s to 1.33s), since injection/stacking and acceleration can now be performed in parallel. [1]

Slip stacking is a process in which 6 batches of 8 GeV protons from the Fermilab Booster are 'stacked' on top of an additional 6 batches of beam. It has been performed and studied previously in the Main Injector [2], using an RF system which was designed primarily for the acceleration of beam. The process is as follows: Six Booster batches are injected into the Recycler Ring and captured by one RF system at 52.809 MHz, and then decelerated slightly by changing the frequency of the RF system by 1260 Hz. Then, six more batches are injected and captured by a second RF system operating at 52.809 MHz. Since the two sets of batches have slightly different energies, one set 'slips' azimuthally with respect to the other. When the two sets of batches are aligned, they are injected into the Main Injector, captured, and then accelerated to 120 GeV.

## CAVITY DESIGN PARAMETERS

Slip stacking requires two cavities with $f_0$ = 52.809 MHz, a maximum $V_{peak}$= 150 kV and approximately 10 kHz of tuning range. The cavities must be tuneable by 1260 Hz for slip stacking itself, and must also have compensation for frequency drift due to heating and variation of cooling water temperature (~ 0.5 kHz/°F). The maximum duty factor will be 60% with a pulse length of 0.8 s.

In the interest of minimizing beam loading, the cavity dimensions were chosen to minimize R/Q while taking into account other practical constraints, such as being able to fit in the allowed area in the tunnel.

The cavity geometry is shown in Figure 1. This is a quarter wave coaxial resonator. The R/Q of 13 Ω is substantially smaller than that in the Main Injector cavities (103 Ω). This is due mainly to the large ratio of diameters of the outer and center conductors. Other key parameters are shown in Table 1.

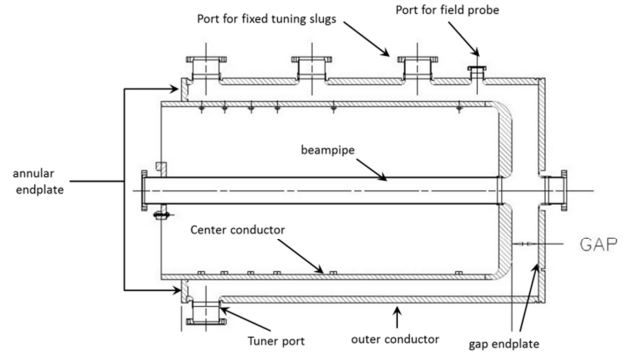

Figure 1: Cavity Geometry and Features

Table 1: Cavity Design Parameters

| | |
|---|---|
| $f_0$ | 52.809 MHz |
| $Z_0$ | 10.2 Ω |
| Vpeak (max) | 150 kV |
| Maximum power | 150 kW |
| Rshunt | 75 kΩ |
| Q | 5800 |
| R/Q | 13 Ω |
| Outer conductor ID | 32 in |
| Inner conductor OD | 27 in |
| nominal gap width* | 2.9 in |
| Inner conductor length** | 49.75 in |
| Step up ratio | 6 |

*Based on pre-weld dimensions. Varied between cavities.
**Nominal length from cavity short to gap end

The initial design parameters and calculations were checked by constructing a model cavity from copper sheet metal [3]. Obviously this did not allow for vacuum or any high power tests, however, it was very useful for

---

verifying and/or measuring key quantities at low power. These include factors such as resonant frequency dependence on various dimensions, higher order modes, and fast tuner development.

### Input Coupler

The cavity input coupler is a loop type coupler which connects to a 9-3/16″, 50 ohm transmission line through a copper and stainless steel bellows. The vacuum window is constructed from a conical 94% alumina ceramic brazed to copper rings. The rings are welded to additional copper sections to form the coupler outer and inner conductors. The coupler 2″ diameter inner conductor is welded to the cavity inner conductor 7.5″ from the shorted end, and the coupler outer conductor is welded to the cavity outer conductor. This results in a cavity accelerating gap to input coupler voltage step-up ratio of 6.

## CAVITY CONSTRUCTION

The cavity center and outer conductors were forged at Scot Forge, Spring Grove, IL from OFHC copper ingots. Following the forging they were machined to final dimensions and ports were added. The circular and annular regions which form the ends of the cavity were also machined from solid blocks of OFHC copper.

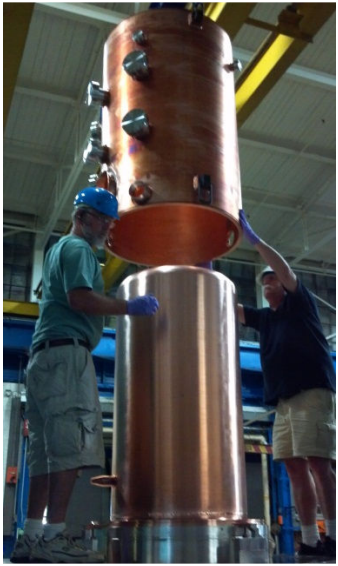

Figure 2: Positioning of outer conductor before electron beam welding.

The cavity center conductor, outer conductor, annual and circular end sections were electron beam welded at Sciaky Inc., Chicago, IL. Figure 2 shows the outer conductor being placed over the center conductor, after the annular end piece was welded to the center conductor. The cavity frequency was tuned after welding the center conductor to the annular ring by adjusting (by machining) the thickness of the endplate near the gap. This was necessary since the cavity frequency changed from pre- to post-weld, presumably because of changes in the effective dimensions due to the welding. The frequency dependence on gap width was measured to be 1.2±0.2 kHz/mil.

In order to maintain a uniform temperature at high duty factor, there are 12 cooling lines each on the cavity center and outer conductors. Each set has a flow rate of 60 gal/min of 95 °F cooling water. The outer cooling lines and were brazed to the cavity. The lines on the ID of the center conductor were shrink-fit. At completion, each cavity weighs 4500 lbs. Some additional aspects of the cavity mechanical design are discussed in [4].

### Cavity Ports

The cavity outer conductor has eight 3.87″ diameter ("larger") and four 1.65″ diameter ("smaller") ports. Furnace brazed copper to stainless steel transitions are used to connect either 6″ or 2-3/4″ Conflat® flanges (depending on the port size) to the copper cavity body. One exception is one of the larger ports used for a fast tuner, which uses a copper flange and not a Conflat.

Three of the larger ports located 35″ (on center) from the cavity shorted end were used for fixed tuning slugs. The slugs are OFHC copper cylinders with a diameter of 3.8″ and a 0.5″ radius on the cavity end. They are attached to water cooled Conflat blanks and protrude nominally 1″ into the cavity. Final machining of the length of the tuning slugs sets the cavity center frequency to 52.809 MHz.

The four smaller ports are all located near the cavity gap end: 44″ from the cavity short. Two of these contain disk shaped capacitively coupled field probes for measurement of cavity gap voltage. These were calibrated by simultaneously measuring the field probe power and the cavity gap voltage with an Agilent 85024A High Frequency Probe and network analyzer. The voltage ratio is nominally $10^5$:1.

A third smaller port contains a 1″ diameter copper gap shorting slug which can be used in the case where there is a third inactive cavity installed in the ring. The slug is mounted on a linear vacuum feedthrough, which can be remotely moved into the cavity to contact the center conductor.

In one of the larger ports located 3″ from the cavity shorted end and opposite the input coupling loop, a fast tuner is loop-coupled to the cavity through a 2.5″ diameter coaxial ceramic window. The loop area is adjusted to give a nominal coupling impedance of 50 Ω. The tuner is a half-wavelength 50 ohm transmission line which is partially loaded with Al doped YIG and shorted at one end. The garnet section is immersed in a variable solenoidal magnetic field which provides for fast cavity tuning. The fast tuner, which has a range of 10 kHz, is described in more detail in a separate contribution to these proceedings [5].

## RF POWER

The Fermilab Tevatron has been decommissioned and we were thus able to reuse the Tevatron RF Stations; this includes the power amplifiers, modulators, 9-3/16″ transmission line, and impedance matching anode resonators. The Tevatron RF systems used essentially the same power amplifiers as does the Main Injector, and these are described in [6].

Power is provided by a cathode driven Eimac Y567B (4CW150000E) 150kW power tetrode. The tetrode is coupled with a monolithic ceramic blocking capacitor to a cylindrical aluminium quarter wave resonator which is used to match the power amplifier output impedance into a 9-3/16″, 50 ohm transmission line. This tap point is at one half the peak voltage. The transmission line includes a variable length phase shifter and a 3:1 coaxial transformer with $3\lambda/2$ between the transformer voltage maximum and the cavity input. Taking into account the cavity input coupler step-up ratio of 6, the 3:1 transformer, and the half voltage tap point on the anode resonator, the total cavity step up ratio from the plate of the tetrode to the cavity gap is 9.

## HIGHER ORDER MODES

Both the measured and calculated lowest frequency higher order modes of the cavity are shown in Table 2. Modes are classified as TEM or TE. The TEM modes are of more concern as they can be easily excited by the beam. All TM modes have frequencies of above ~2 GHz. The 3$^{rd}$ harmonic TEM mode is damped by a ½″ Heliax® quarter wave coaxial line with a capacitive pickup on the 9-3/16" transmission line (near a voltage maximum). For damping, a 5 W, 50 Ω load is connected to the heliax line near the 50 Ω tap point.

For the modes listed in Table 2, the frequencies of the TE modes are calculated analytically for a coaxial waveguide, assuming a line length of 50″, and using $\lambda=\lambda_g/(1+(\lambda_g/\lambda_c)^2)^{1/2}$, where $\lambda_g$ and $\lambda_c$ are the guide and cutoff wavelengths of the coaxial line [7]. The TEM modes were calculated using the LANL Superfish program [8].

Table 2: Higher Order Modes

| Calculated f (MHz) | Measured f (MHz) | Type | order | Calculated f (MHz) | Measured f (MHz) | Type | order |
|---|---|---|---|---|---|---|---|
| 52.809 | 52.809 | TEM | $\lambda/4$ | 385.8 | 388 | TE$_{31}$ | $\lambda_g/4$ |
| 140.36 | 151.72 | TE$_{11}$ | $\lambda_g/4$ | 389.79 | 392.44 | TE$_{21}$ | $5\lambda_g/4$ |
|  | 154.695 |  |  |  | 393.085 |  |  |
| 154.748 | 157.780 | TEM | $3\lambda/4$ | 421 | 428.15 | TE$_{31}$ | $3\lambda_g/4$ |
| 218.081 | 222.19 | TE$_{11}$ | $3\lambda_g/4$ |  | 428.870 |  |  |
|  | 226.13 |  |  | 429.882 | 437.130 | TEM | $9\lambda/4$ |
| 245.228 | 247.17 | TEM | $5\lambda/4$ | 432.281 | 432.42 | TE$_{11}$ | $7\lambda_g/4$ |
| 261.45 | 267.795 | TE$_{21}$ | $\lambda_g/4$ | 482.74 | 470.9 | TE$_{31}$ | $5\lambda_g/4$ |
|  | 271.630 |  |  |  | 473.165 |  |  |
| 310.186 | 308.890 | TE$_{21}$ | $3\lambda_g/4$ | 485.32 | 475.13 | TE$_{21}$ | $7\lambda_g/4$ |
|  | 312 |  |  |  | 486.66 |  |  |
| 318.485 | 318.63 | TE$_{11}$ | $5\lambda_g/4$ | 512.8 | 516.635 | TE$_{41}$ | $\lambda_g/4$ |
|  | 323.81 |  |  |  | 518.15 |  |  |
| 331.111 | 337.80 | TEM | $7\lambda/4$ |  |  |  |  |

## CAVITY TESTING, INSTALLATION, AND COMMISSIONING

Prior to their installation in the tunnel, two cavities were tested to full power in a cavity test cave. The cavities were first subjected to low power (up to 500 W) CW RF for one week, to condition away multipacting between the outer and inner conductors. In this case the cavity was driven through the fast tuner port (with the tuner disconnected). A photo of the cavities installed in the Main Injector tunnel is shown in Figure 3.

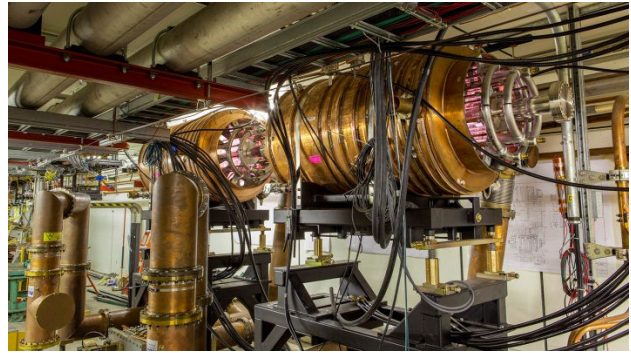

Figure 3: Two complete cavities installed in the Recycler.

## ACKNOWLEDGMENTS

We would like to acknowledge the huge efforts of the Accelerator Division Mechanical Support and RF Groups, and everyone involved in this project. We would also like to thank Marty Murphy for the photograph in Figure 3.